\documentclass[vecphys]{svmult}

\usepackage{axodraw}

\usepackage{makeidx}         
\usepackage{graphicx}        
\usepackage{multicol}        
\usepackage[bottom]{footmisc}

\makeindex             

\newcommand{\be}{\begin{equation}}
\newcommand{\ee}{\end{equation}}

\newcommand{\bea}{\begin{eqnarray}}
\newcommand{\eea}{\end{eqnarray}}

\def\Fig#1{Fig.~\ref{#1}}

\def\tlambda{{\tilde\lambda}}

\newbox\SlashedBox
\def\slashed#1{\setbox\SlashedBox=\hbox{#1}
\hbox to 0pt{\hbox to 1\wd\SlashedBox{\hfil/\hfil}\hss}#1}
\def\hboxtosizeof#1#2{\setbox\SlashedBox=\hbox{#1}
\hbox to 1\wd\SlashedBox{#2}}

\newbox\charbox
\newbox\slabox
\def\s#1{{      
        \setbox\charbox=\hbox{$#1$}
        \setbox\slabox=\hbox{$/$}
        \dimen\charbox=\ht\slabox
        \advance\dimen\charbox by -\dp\slabox
        \advance\dimen\charbox by -\ht\charbox
        \advance\dimen\charbox by \dp\charbox
        \divide\dimen\charbox by 2
        \raise-\dimen\charbox\hbox to \wd\charbox{\hss/\hss}
        \llap{$#1$}
}}

\def\spa#1.#2{\langle#1\,#2\rangle}
\def\spb#1.#2{[#1\,#2]}

\def\spab#1.#2.#3{\langle\mskip-1mu{#1}
                  | #2 | {#3}]}

\def\spba#1.#2.#3{[\mskip-1mu{#1}
                  | #2 | {#3}\rangle}

\def\spbb#1.#2.#3.#4{[\mskip-1mu{#1}
                     | {#2} \ {#3} | {#4}]}

\def\spaa#1.#2.#3.#4{\langle\mskip-1mu{#1}
                     | {#2} \ {#3} | {#4}\rangle}

\begin{document}

\title*{Twistors and Unitarity}
\author{Pierpaolo Mastrolia
}
\institute{Institute of Theoretical Physics, University of Z\"urich,
CH-8057 Switzerland
}
%
%
\maketitle


\vspace*{-1.0cm}

{\bf Introduction.}
Motivated by the demand for a higher accuracy
description of the multi-particle processes that will take
place at the upcoming LHC experiment, 
the theoretical efforts to develop new computational techniques 
for perturbative calculation
have received a strong boost, 
stemmed from {\it on-shell} techniques and unitarity-based methods.  \\
{\bf Spinors and Twistors.}
\def\tlambda{{\widetilde\lambda}}
The spinor-helicity representation of QCD amplitudes, 
that was developed in the 1980s, has been
an invaluable tool in perturbative computations ever since.  
Accordingly, one can express any on-shell massless vector as the
tensor product of two spinors 
carrying different flavour under $SU(2)$, 
say $\lambda$ and $\tlambda.$
Instead of Lorentz inner products of momenta, 
amplitudes can be expressed in terms of spinor products.
The tree-level amplitudes
for the scattering of $n$ gluons vanish, when  
the helicities of the gluons 
are either {\it a)} all the same, or {\it b)} all the same, but one of 
opposite helicity.
The first sequence of nonvanishing tree amplitudes is called 
maximally helicity-violating (MHV), formed by 
two gluons of negative helicity, and the rest of
positive helicity ones, whose simple form \cite{ParkeTaylor,BGrecrel} 
involves only spinor products of a single flavour, $\lambda$. 
Witten observed that under the twistor transform \cite{Penrose}, 
MHV amplitudes, due to their {\it holomorphic} character, appear to be 
supported on lines in twistor space, thus revealing an 
intrinsic structure of local objects, once 
transformed back to Minkowski space \cite{Witten}. 
Amplitudes with more negative-helicity
gluons are represented by 
intersecting segments \cite{RSV,CSW}, each representing a MHV arrangement 
of particles. \\
{\bf CSW Construction.}
\label{ssec:CSW}
\index{CSW Formalism}
This representation turned out in a novel diagrammatic interpretation,
in the form of the Cachazo--Svr\v{c}ek--Witten (CSW) construction~\cite{CSW},
which can be used alternatively to Feynman rules.
To compute an amplitude for $n$ gluons, out of which 
$m$ gluons carry negative helicity, one
has to:
draw all the possible graphs connecting the $n$ external
gluons, with at least three legs in each vertex (node);
assign the helicities, $\pm$, to the internal legs;
keep the graphs having only MHV-configuration nodes;
consider any internal line as a scalar propagator;
sum over all and only MHV-diagrams.
The prescription for the spinor products involving an off-shell 
momenta is realized by the introduction of a 
light-cone {\it reference} spinor \cite{CSW,BBK}
which disappears after adding all the contributions up.
Because of the efficiency of the MHV rules for gluon tree amplitudes in QCD, 
they were soon generalized to other processes, like
scattering of massless 
fermions~\cite{ExtFerm},
amplitudes with a Higgs boson~\cite{ExtHiggs},
and more general objects carrying a Lorentz index, like the fermionic 
currents, to compute amplitudes involving electroweak vector 
bosons~\cite{ExtVector}. 

\noindent
{\bf BCFW On-Shell Recurrence Relation.}
\label{ssec:BCFW}
\index{BCFW Recurrence Relation}
In parallel with the extension of tree-level MHV rules to several 
processes, the twistor structure of one-loop amplitudes
began to be investigated - a long list of referencs should be added here!
Unexpectedly, these one-loop computations led to a new, more compact 
representation of tree-amplitudes (appearing in the 
infrared-divergent parts of the one-loop amplitudes) 
\cite{RSVNewTree}.
%
\begin{figure}[t]
\vspace*{0.7cm}
$$
{\small
\begin{picture}(0,0)(0,0)
\SetScale{0.7}
\SetWidth{1.0}
\Line(0, 0)(-10, -35) \Text(-10,-30)[]{{$1$}}
\Line(0, 0)(-30, -25) \Text(-25,-20)[]{{$2$}}
\Line(0, 0)(-30,  25) \Text(-35, 20)[]{{$k-1$}}
\Line(0, 0)(-10,  35) \Text(-10, 30)[]{{$k$}}
\Line(0, 0)(10,  35) \Text(15, 30)[]{{$k+1$}}
\Line(0, 0)(30,  25) \Text(35, 20)[]{{$k+2$}}
\Line(0, 0)(30,  -25) \Text(35, -20)[]{{$n-1$}}
\Line(0, 0)(10,  -35) \Text(10, -30)[]{{$n$}}
\GOval(0,0)(20,20)(0){.9} 
\Text(0, 0)[]{{$A_n$}}
\Vertex(-25,-10){1}
\Vertex(-27,  0){1}
\Vertex(-25,+10){1}
\Vertex( 25,-10){1}
\Vertex( 27,  0){1}
\Vertex( 25,+10){1}
\end{picture}
%
\hspace*{2cm}
=
\quad \sum_{k,h}
\hspace*{2cm}
%
\begin{picture}(50,0)(0,0)
\SetScale{0.7}
\SetWidth{1.0}
\Line(-17, 0)(55, 0) \Text(0,5)[]{{$-h$}} \Text(17,-5)[]{{$h$}}
\Line(-33, 0)(-45, -35) \Text(-35,-30)[]{{$\hat{1}$}}
\Line(-33, 0)(-65, -25) \Text(-50,-20)[]{{$2$}}
\Line(-33, 0)(-65,  25) \Text(-55, 22)[]{{$k-1$}}
\Line(-33, 0)(-45,  35) \Text(-35, 30)[]{{$k$}}
\Line(55, 0)(65,  35) \Text(50, 30)[]{{$k+1$}}
\Line(55, 0)(85,  25) \Text(75, 20)[]{{$k+2$}}
\Line(55, 0)(85, -25) \Text(75, -20)[]{{$n-1$}}
\Line(55, 0)(65, -35) \Text(50, -30)[]{{$\hat{n}$}}
\GOval(-33,0)(15,20)(0){.9}
\GOval(55,0)(15,20)(0){.9}
\Text(-25,0)[]{{\tiny{$A_{k+1}$}}} 
\Text(38,0)[]{{\tiny{$A_{n-k+1}$}}}
\Vertex(-60,-10){1}
\Vertex(-63,  0){1}
\Vertex(-60,+10){1}
\Vertex( 80,-10){1}
\Vertex( 83,  0){1}
\Vertex( 80,+10){1}
\end{picture}
} 
$$
\vspace*{0.5cm}
\caption{BCFW Recurrence Relation for tree amplitudes.}
\label{fig:BCFW}
\vspace*{-0.5cm}
\end{figure}
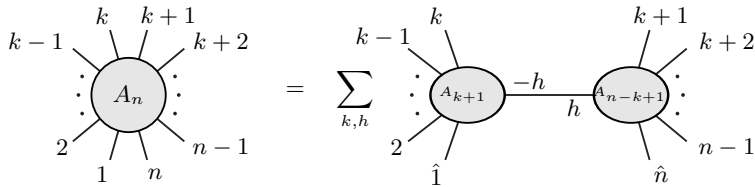
%
These results could be reinterpreted
as coming from a quadratic recursion, the so called
BCFW on-shell recurrence relation~\cite{BCFRecursion,BCFW},
depicted diagrammatically in \Fig{fig:BCFW}.
The amplitude is represented as a sum of products of lower-point 
amplitudes, evaluated on shell, but for {\it complex} values of 
the shifted momenta (denoted by a hat).  
There are two sums.
The first is over the helicity $h$ of an internal gluon 
propagating between the two amplitudes.
The second sum is over an
integer $k$, which labels the 
partitions of the set $\{1,2,\ldots,n\}$
into two consecutive subsets (with minimum 3 elements), 
where the labels of the shifted momenta, say
$1$ and $n$, belong to distinct subsets. 
Britto, Cachazo, Feng and Witten (BCFW) showed that
these on-shell recursion relations have a very general
proof, relying only on factorization and complex analysis~\cite{BCFW}.  
Accordingly, the same
approach can be applied to amplitudes in General Relativity~\cite{Gravity},
to scattering with massive theories~\cite{Massive},
and to the rational coefficients of the integrals that
appear in (special helicity configuration) 
one-loop amplitudes~\cite{IntegralCoefficients}.  
In the contest of the on-shell formalism, the CSW construction 
has been interpreted as a variant of the BCFW relation, 
obtained by a very peculiar analytic continuation of 
momenta~\cite{Risager}. \\
%
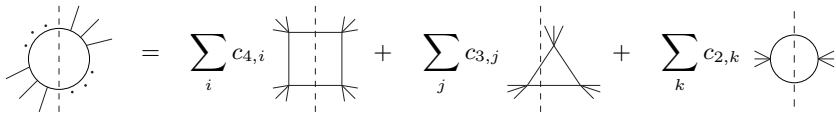
\begin{figure*}[h]
$$
\begin{picture}(0,0)(0,0)
\SetScale{0.5}
\SetWidth{0.5}
\Line(15,40)(5,10)    
\Line(30, 30)(10, 10) 
\Line(40,15)(0,5)     
\Line(-40,-15)(0,5)     
\Line(-30,-30)(-10,-10) 
\Line(-15,-40)(-5,-10)  
\GOval(0,0)(23,23)(0){1.0}
\Vertex(-25,+10){1}
\Vertex(-20,+20){1}
\Vertex(-10,+25){1}
\Vertex(+25,-10){1}
\Vertex(+20,-20){1}
\Vertex(+10,-25){1}
\DashLine(0,40)(0,-40){5}
\end{picture}
\hspace*{1.0cm} 
=
\hspace*{0.3cm} 
\sum_i c_{4,i} \qquad 
\begin{picture}(0,0)(0,0)
\SetScale{0.5}
\SetWidth{0.5}
\Line(-20,-20)(20,-20)
\Line(20,-20)(20,20)
\Line(20,20)(-20,20)
\Line(-20,20)(-20,-20)
\Line(-20,-20)(-32,-22)
\Line(-20,-20)(-30,-30)
\Line(-20,-20)(-22,-32)
\Line(20,-20)(32,-22)
\Line(20,-20)(30,-30)
\Line(20,-20)(22,-32)
\Line(20,20)(32,22)
\Line(20,20)(30,30)
\Line(20,20)(22,32)
\Line(-20,20)(-32,22)
\Line(-20,20)(-30,30)
\Line(-20,20)(-22,32)
\DashLine(0,40)(0,-40){5}
\end{picture}
%
\hspace*{0.7cm} 
+
\quad
\sum_j c_{3,j}
\hspace*{0.7cm} 
%
\begin{picture}(0,0)(0,0)
\SetScale{0.5}
\SetWidth{0.5}
\Line(-20,-20)(0,10)
\Line(20,-20)(0,10)
\Line(20,-20)(-20,-20)
\Line(0,10)(-7,25)
\Line(0,10)( 0,25)
\Line(0,10)( 7,25)
\Line(-20,-20)(-30,-30)
\Line(-20,-20)(-35,-25)
\Line(-20,-20)(-35,-20)
\Line(20,-20)(30,-30)
\Line(20,-20)(35,-25)
\Line(20,-20)(35,-20)
\DashLine(-10,40)(-10,-40){5}
\end{picture}
%
\hspace*{0.7cm} 
+
\quad
\sum_k c_{2,k}
\hspace*{0.7cm} 
%
\begin{picture}(0,0)(0,0)
\SetScale{0.6}
\SetWidth{0.6}
\GOval(0,0)(15,15)(0){1}
\Line(-15,0)(-25,+5)
\Line(-15,0)(-25,0)
\Line(-15,0)(-25,-5)
\Line( 15,0)(25,+5)
\Line( 15,0)(25,0)
\Line( 15,0)(25,-5)
\DashLine(0,30)(0,-30){5}
\end{picture}
%
$$
\caption{Double-cut of 
a one-loop amplitude in
  terms of the master-cuts.}
\label{Deco}
\vspace*{-0.5cm}
\end{figure*}
%
{\bf Unitarity-based Methods.}
\label{ssec:UBM}
\index{Unitarity methods}
Unitarity-based methods are an effective
tool to compute amplitudes at loop-level 
\cite{BernZX}--\cite{UnitarityMethod}.
The analytic expression of any one-loop amplitude, 
written, by Passarino-Veltman reduction, in terms of a basis of scalar 
integral functions (boxes, triangles, and bubbles),
may contain 
a polylogarithmic structure and a pure rational term.
To compute the amplitude, it is sufficient to compute
the (rational) coefficients of the linear combination separately, 
and the principle of the unitarity-based method is to exploit the unitarity
cuts of the scalar integrals to extract their coefficients,
see \Fig{Deco}.
Unitarity in four-dimension (4D), 
requiring the knowledge of the four-dimensional 
on-shell tree amplitudes which are
sewed in the cut, is sufficient to compute the polylogarithmic terms 
and the trascendenal constants of one-loop amplitudes. 
By exploiting the analytic continuation of tree-amplitudes to complex 
spinors, and the properties of the complex integration,
new techniques have generalized the cutting rules.
On the one side, the {\it quadruple-cut} technique \cite{GenUnit} 
yields the immediate computation of boxes' coefficient.
On the other side, the polylogarithmic structure related to 
box-, triangle- and 
bubble-functions can be detected by a {\it double-cut} 
and computed by a 
novel way of performing the cut-integral \cite{DoubleCuts,BFM},
which reduces the integration to the extraction of residues in spinor 
variables.
However, on general grounds, amplitudes in nonsupersymmetric theories, 
like QCD, suffer of rational ambiguities
that are not detected by the four-dimensional dispersive integrals,
and $D$-dimensional unitarity \cite{BernMorgan,SelfDual,BSTGenUn,
Anastasiou:2006jv} can be used to determine these terms as well.
Alternatively, 
according to the combined {\it unitarity-bootstrap} approach, 
after computing the cut-containing terms 
by 4D-unitarity, the use of 
a BCFW-like recurrence relation
yields the reconstruction of the rational part
\cite{OneLoopRecursion}.
%
In the very recent past - after the IFAE 2006 workshop -
an optimized tool has been developed by 
tailoring the Passarino-Veltman reduction on the integrals that are 
responsible of the rational part of scattering amplitudes
\cite{Xiao:2006vr}.
That has given rise to further refinements and new developments 
of algorithms for the tensor reduction of Feynman integrals 
\cite{Ossola:2006us,Binoth:2006hk}.
I conclude with the Tab. 1, which collects the efforts of the last 12
years,
invested in the analytic computation of the six-gluon amplitude in QCD, 
numerically evaluated not so long ago \cite{Ellis:2006ss}.

\begin{table}[t]
\begin{center}
\begin{tabular}{|c||c|c|c|c|}
\hline
Amplitude  &${\cal N}=4$& 
            ${\cal N}=1$ & 
            ${\cal N}^{\rm cut}=0$ &
            ${\cal N}^{\rm rat}=0$ \\
\hline
\hline
$- - + + + +$ 
           & \cite{BernZX}
           & \cite{BernCG}
           & \cite{BernCG}
           & \cite{BernCQ}
\\
\hline
$- + - + + +$ 
           & \cite{BernZX}
           & \cite{BernCG}
           & \cite{BedfordNH}
           & \cite{Berger:2006vq,Xiao:2006vr}
\\
\hline
$- + + - + +$ 
           & \cite{BernZX}
           & \cite{BernCG}
           & \cite{BedfordNH}
           & \cite{Berger:2006vq,Xiao:2006vr}
\\
\hline
$- - - + + +$
           & \cite{BernCG}
           & \cite{BidderTX}
           & \cite{IntegralCoefficients,BFM}
           & \cite{Berger:2006ci}
\\
\hline
$- - + - + +$ 
           & \cite{BernCG}
           & \cite{DoubleCuts,BidderVX,BidderRI}
           & \cite{BFM}
           & \cite{Xiao:2006vr}
\\
\hline
$- + - + - +$ 
           & \cite{BernCG}
           & \cite{DoubleCuts,BidderVX,BidderRI}
           & \cite{BFM}
           & \cite{Xiao:2006vr}
\\
\hline
\end{tabular}
\end{center}
\caption{The analytic computation of  
the one-loop six-gluon amplitude in QCD}
\label{table:status6}
\end{table}

\end{document}